\documentstyle[art10]{article}
\begin{document}

\begin{titlepage}

\title{Octonion XY-Product }

\author{Geoffrey Dixon\thanks{supported in part by viewers like
you.} \\
Department of Mathematics or Physics \\
 Brandeis University \\
Waltham, MA 02254 \\
 email: dixon@binah.cc.brandeis.edu \\
\and Department of Mathematics \\
University of Massachusetts \\
Boston, MA 02125 \\
 email: dixon@umbsky.cc.umb.edu}

\maketitle

\begin{abstract} The octonion X-product changes the octonion
multiplication table, but does not change the role of the identity.
The XY-product is very similar, but shifts the identity as well.
This will be of interest to those applying the octonions to string
theory.
\end{abstract}

\end{titlepage}

\section*{1. Moufang Identities.}

The Moufang identities are listed here for future reference.  I list
them in their conventional form and in the form I introduced in {\bf
[1]}. \\
%
%
\begin{equation}
\fbox{$
\begin{array}{ll} (xy)(ax)=x(ay)x; & (xy)_{L}x_{R} = x_{L}x_{R}y_{L};
\\ \\ (xa)(yx)=x(ya)x; & (yx)_{R}x_{L} = x_{L}x_{R}y_{R}; \\ \\
(xax)y=x(a(xy)); & y_{R}x_{L}x_{R} = x_{L}(xy)_{R}; \\ \\
y(xax)=((yx)a)x; & y_{L}x_{L}x_{R} = x_{R}(yx)_{L}. \\
\end{array}
$}
\end{equation}

\section*{2 ${\bf O}_{X,Y}$.}

The X-product {\bf [1][2]} changes {\bf O} to ${\bf O}_{X}$, which is
isomorphic to {\bf O}.  Furthermore, the identities of {\bf O} and
${\bf O}_{X}$ are both $e_{0}=1$.  However, it is possible to modify
the octonion product in such a way that $e_{0}$ is not the identity
of the result.  In particular, define \\
%
%
\begin{equation}
\fbox{$
\begin{array}{cl} A{\circ_{X,Y}}B & = (AX)(Y^{\dagger}B) \\ \\ & =
A{\circ_{X}}((XY^{\dagger}){\circ_{X}}B) \\ \\ & =
(A{\circ_{Y}}(XY^{\dagger})){\circ_{Y}}B, \\
\end{array}
$}
\end{equation} \\
where as usual we assume that both $X,Y \in
S^{7}$ (the X-product is obtained by setting $X=Y$).  Let ${\bf
O}_{X,Y}$ be {\bf O} with this modified product.

The question is, is this still isomorphic {\bf O} itself?  Let ${\bf
O}$ be the copy of the octonions employing the cyclic multiplication
table introduced in {\bf [1]}, and let
$$
E_{a}, \; a=0,...,7,
$$
be a basis for ${\bf O}_{X,Y}$.  We will attempt to make
assignments for the $E_{a}$ so that they satisfy the same table, and
consequently
$e_{a}
\longrightarrow E_{a}, \; a=0,...,7,$ will be an isomorphism
from  {\bf O} to ${\bf O}_{X,Y}$.  \\ \\
IDENTITY \\

As a start, it is not difficult to prove that in general,
%
%
\begin{equation}
\fbox{$  E_{0} = YX^{\dagger}
$}
\end{equation} (I leave it to the reader to establish that this is a
two-sided identity). \\ \\
HERMITIAN CONJUGATION \\

Hermitian conjugation is also altered, and it will help to determine
this before proceeding.  In particular,  let $A^{\star}$ denote the
${\bf O}_{X,Y}$ conjugate of $A$.  It must satisfy
%
%
\begin{equation} (AX)(Y^{\dagger}A^{\star}) =
(A^{\star}X)(Y^{\dagger}A) = \|A\|^{2}E_{0} = AA^{\dagger}E_{0},
\end{equation}
where $\|A\|^{2}$ is the square of the norm of $A$.
Therefore, (4) implies the two results,
%
%
\begin{equation}
\fbox{$
\begin{array}{l} A^{\star} = \|A\|^{2}Y((AX)^{-1}E_{0}) =
Y((X^{\dagger}A^{\dagger})(YX^{\dagger})) =
Y(X^{\dagger}(A^{\dagger}Y)X^{\dagger}), \\ \\ A^{\star} =
\|A\|^{2}(E_{0}(Y^{\dagger}A)^{-1})X^{\dagger} =
((YX^{\dagger})(A^{\dagger}Y))X^{\dagger} =
(Y(X^{\dagger}A^{\dagger})Y)X^{\dagger} \\
\end{array}
$}
\end{equation} \\
(the Moufang identities were used in (5), along
with $A^{-1} = A^{\dagger}/\|A\|^{2}$).  Proving the two versions of
$A^{\star}$ in (5) are the same is equivalent to proving
$$
Y_{L}X_{L}^{\dagger}X_{R}^{\dagger}Y_{R} =
X_{R}^{\dagger}Y_{L}Y_{R}X_{L}^{\dagger}.
$$
I leave it to the reader to prove this using the righthand
identities in (1). \\ \\
ISOMORPHISM \\

In order to prove the general isomorphism of {\bf O} amd ${\bf
O}_{X,Y}$, we will start with the simpler case of ${\bf O}_{1,Z}$,
the product of which is
%
%
\begin{equation}
A{\circ_{1,Z}}B = A(Z^{\dagger}B).
\end{equation}
Therefore,
%
%
\begin{equation}
E_{0} = Z,
\end{equation}
and
%
%
\begin{equation}
A^{\star} = ZA^{\dagger}Z.
\end{equation} \\

Without any loss in generality we can set
%
%
\begin{equation}
Z=Z^{0}+Z^{7}e_{7}, \; \; \; \; ZZ^{\dagger} =
(Z^{0})^{2}+(Z^{7})^{2} = 1.
\end{equation}
After playing around a bit, I came up with the
following assignments:
%
%
\begin{equation}
\begin{array}{c}
E_{1} = Ze_{1} = e_{1}Z^{\dagger}; \; \; \;  \;
E_{5} = Ze_{5} = e_{5}Z^{\dagger}; \\ \\ E_{2} = Ze_{2} =
e_{2}Z^{\dagger}; \; \; \;  \; E_{3} = Ze_{3} = e_{3}Z^{\dagger}; \\
\\ E_{4} = Z^{\dagger}e_{4} = e_{4}Z; \; \; \;  \; E_{6} =
Z^{\dagger}e_{6} = e_{6}Z; \\ \\ E_{7} = Ze_{7} = e_{7}Z. \\
\end{array}
\end{equation}
Observe that
%
%
\begin{equation}
\begin{array}{c}
(Ze_{a})^{\star} = -Ze_{a}Z^{\dagger}Z = -Ze_{a},
\\ \\ (e_{a}Z)^{\star} = -ZZ^{\dagger}e_{a}Z = -e_{a}Z \\
\end{array}
\end{equation}
(no parentheses needed), so each of the $E_{a}, \;
a=1,...,7$, is perpendicular to $E_{0}=Z$. \\

To make life easier, we'll first check that these elements
anticommute.  That is, if
$a,b\in\{1,...,7\}$ are distinct, then
%
%
\begin{equation} E_{a}{\circ_{1,Z}}E_{b} = -E_{b}{\circ_{1,Z}}E_{a}.
\end{equation} \\
\begin{flushleft} PROOF OF ANTICOMMUTATION \\
\end{flushleft} CASE I: $E_{a}=e_{a}Z$, $E_{b}=e_{b}Z$. \\

Therefore $a,b\in\{4,6,7\}$, a quaternionic triple.  Since $Z$ is
also linear in $1$ and $e_{7}$, the parentheses below can be dropped.
%
%
\begin{equation}
\begin{array}{cl}
E_{a}{\circ_{1,Z}}E_{b} & =
(e_{a}Z)(Z^{\dagger}(e_{b}Z)) \\ \\ & = e_{a}ZZ^{\dagger}e_{b}Z \\ \\
& = e_{a}e_{b}Z \\ \\ & = -e_{b}e_{a}Z \\ \\ & =
-E_{b}{\circ_{1,Z}}E_{a}.
\end{array}
\end{equation}
CASE II: $E_{a}=Ze_{a}$, $E_{b}=e_{b}Z$. \\

In this case,
%
%
\begin{equation}
a\in\{1,2,3,5\}, \; \; b\in\{4,6,7\}.
\end{equation}
This is used twice below.
%
%
\begin{equation}
\begin{array}{cll}
E_{a}{\circ_{1,Z}}E_{b} & =
(Ze_{a})(Z^{\dagger}(e_{b}Z)) & \\ \\  & =
(Ze_{a})(Z^{\dagger}e_{b}Z) & \mbox{Associate} \\ \\  & =
Z[e_{a}(Z^{\dagger}e_{b})]Z & \mbox{Moufang} \\ \\  & =
[e_{a}(Z^{\dagger}e_{b})]Z^{\dagger}Z & \mbox{From 9,14} \\ \\  & =
e_{a}(Z^{\dagger}e_{b}) &  \\ \\  & = [(e_{b}Z)e_{a}]^{\dagger} &  \\
\\  & = -(e_{b}Z)e_{a} & \mbox{From 9,14} \\ \\  & =
(e_{b}Z)(Z^{\dagger}(Ze_{a})) & \\ \\  & = -E_{b}{\circ_{1,Z}}E_{a}.
& \\
\end{array}
\end{equation} \\
\begin{flushleft}
CASE III: $E_{a}=Ze_{a}$, $E_{b}=Ze_{b}$. \\
\end{flushleft}

In this final case, $a,b\in\{1,2,3,5\}$.
%
%
\begin{equation}
\begin{array}{cll}
E_{a}{\circ_{1,Z}}E_{b} & =
(Ze_{a})(Z^{\dagger}(Ze_{b})) & \\ \\   & = (Ze_{a})e_{b} &  \\ \\
& = -(Ze_{b})e_{a} & \mbox{See {\bf [1]}} \\ \\  & =
-(Ze_{b})(Z^{\dagger}(Ze_{a})) & \\ \\   & =
-E_{b}{\circ_{1,Z}}E_{a}. & \\
\end{array}
\end{equation}
So (12) is proven. \\ \\
MULTIPLICATION TABLE \\

Because of (12), to complete the multiplication table (and prove the
isomorphism) we need merely check that
%
%
\begin{equation}
\mbox{if }\; \; e_{a}e_{b} = e_{c}, \mbox{ then }\; \;
E_{a}{\circ_{1,Z}}E_{b} = E_{c},
\end{equation}
and
%
%
\begin{equation}
 E_{a}{\circ_{1,Z}}E_{a} = -E_{0}, \; \; a=1,...,7.
\end{equation}

Prove (18) first.  Note that parentheses may be dropped in this
case.  There are two possibilities:
$$
\begin{array}{l} (Ze_{a})(Z^{\dagger}(Ze_{a})) =
Ze_{a}Z^{\dagger}Ze_{a} = Ze_{a}e_{a} = -Z, \\ \\
(e_{a}Z)(Z^{\dagger}(e_{a}Z)) = e_{a}ZZ^{\dagger}e_{a}Z = e_{a}e_{a}Z
= -Z. \\
\end{array}
$$
This proves (18). \\

We'll prove (17) by cases. \\ \\
CASE I: $E_{1}{\circ_{1,Z}}E_{2}$. \\

$$
\begin{array}{cll}
E_{1}{\circ_{1,Z}}E_{2} & =
(Ze_{1})(Z^{\dagger}(Ze_{2})) & \\ \\    & =(Ze_{1})e_{2} &  \\ \\
& = Z^{\dagger}(e_{1}e_{2}) & \mbox{Nonassociativity} \\ \\  & =
Z^{\dagger}e_{6} & \\ \\    & = e_{6}Z & \\ \\ & = E_{6}. & \\
\end{array}
$$
This example covers the four products,
%
%
\begin{equation}
\begin{array}{l}
 E_{1}{\circ_{1,Z}}E_{2} = E_{6}, \; \; \; \; E_{3}{\circ_{1,Z}}E_{5}
= E_{6}, \\ \\ E_{5}{\circ_{1,Z}}E_{2} = E_{4}, \; \; \; \;
E_{1}{\circ_{1,Z}}E_{3} = E_{4}. \\
\end{array}
\end{equation}
CASE II: $E_{a}{\circ_{1,Z}}E_{7}$. \\

In this case again, parentheses may be deleted.
$$
\begin{array}{cll}    E_{a}{\circ_{1,Z}}E_{7} & =
E_{a}(Z^{\dagger}(Ze_{7})) & \\ \\    & = E_{a}e_{7} &  \\ \\      &
= e_{a}Ze_{7} \mbox{ or } Ze_{a}e_{7}  &  \\ \\   & = (e_{a}e_{7})Z
\mbox{ or } Z(e_{a}e_{7}).  &  \\
\end{array}
$$
This example covers the six products,
%
%
\begin{equation}
\begin{array}{l}
 E_{7}{\circ_{1,Z}}E_{1} = E_{5}, \; \; \; \; E_{7}{\circ_{1,Z}}E_{2}
= E_{3},
\; \; \; \; E_{7}{\circ_{1,Z}}E_{4} = E_{6}, \\ \\
E_{5}{\circ_{1,Z}}E_{7} = E_{1}, \; \; \; \; E_{3}{\circ_{1,Z}}E_{7}
= E_{2},
\; \; \; \; E_{6}{\circ_{1,Z}}E_{7} = E_{4}. \\
\end{array}
\end{equation} \newpage

\begin{flushleft}
CASE III: $E_{a}{\circ_{1,Z}}E_{b}$, where
$e_{a}e_{b} = e_{7}$. \\
\end{flushleft}

Parentheses may be deleted in this case too, as all products
associate.  Note that if $E_{a} = Ze_{a}$ (or $e_{a}Z$), then $E_{b}
= Ze_{b}$ (or $e_{b}Z$)).  So
$$
\begin{array}{cll}     E_{a}{\circ_{1,Z}}E_{b} & =
(e_{a}Z)(Z^{\dagger}(e_{b}Z)) & \mbox{or }
(Ze_{a})(Z^{\dagger}(Ze_{b}))\\ \\     & = e_{a}ZZ^{\dagger}e_{b}Z &
\mbox{or }  Ze_{a}Z^{\dagger}Ze_{b} \\ \\ & = e_{a}e_{b}Z & \mbox{or
}  Ze_{a}e_{b} \\ \\ & = e_{7}Z & = Z_{7} \\ \\ & = E_{7}.  &  \\
\end{array}
$$
This example covers the three products,
%
%
\begin{equation}
 E_{1}{\circ_{1,Z}}E_{5} =  E_{2}{\circ_{1,Z}}E_{3} =
 E_{4}{\circ_{1,Z}}E_{6} = E_{7}.
\end{equation} \\
CASE IV: $E_{4}{\circ_{1,Z}}E_{1}$. \\

$$
\begin{array}{cll}      E_{4}{\circ_{1,Z}}E_{1} & =
(e_{4}Z)(Z^{\dagger}(Ze_{1})) &  \\ \\      & = (e_{4}Z)e_{1} &  \\
\\   & = (Z^{\dagger}e_{4})e_{1} &  \\ \\  & = Z(e_{4}e_{1}) &
\mbox{Nonassociativity} \\ \\  & = Ze_{3} &  \\ \\  & = E_{3}.  &  \\
\end{array}
$$
This example covers the four cases,
%
%
\begin{equation}
\begin{array}{l}
 E_{4}{\circ_{1,Z}}E_{1} = E_{3}, \; \; \; \; E_{4}{\circ_{1,Z}}E_{5}
= E_{2}, \\ \\ E_{6}{\circ_{1,Z}}E_{1} = E_{2}, \; \; \; \;
E_{6}{\circ_{1,Z}}E_{3} = E_{5}. \\
\end{array}
\end{equation} \newpage

\begin{flushleft}
CASE V: $E_{2}{\circ_{1,Z}}E_{4}$. \\
\end{flushleft}

Lastly,
$$
\begin{array}{cll}      E_{2}{\circ_{1,Z}}E_{4} & =
(Ze_{2})(Z^{\dagger}(e_{4}Z)) & \\ \\   & =
(Ze_{2})((Z^{\dagger}e_{4})Z)  & \\ \\ & =
Z[e_{2}(Z^{\dagger}e_{4})]Z & \mbox{Moufang} \\ \\ & =
[e_{2}(Z^{\dagger}e_{4})]Z^{\dagger}Z & \mbox{Think about it} \\ \\ &
= e_{2}(e_{4}Z) &  \\ \\ & = (e_{2}e_{4})Z^{\dagger} &
\mbox{Nonassociativity} \\ \\ & = e_{5}Z^{\dagger} &  \\ \\ & =
Ze_{5} &  \\ \\ & = E_{5}. &  \\ \\
\end{array}
$$
This example covers the four products,
%
%
\begin{equation}
\begin{array}{l}
 E_{2}{\circ_{1,Z}}E_{4} = E_{5}, \; \; \; \; E_{3}{\circ_{1,Z}}E_{4}
= E_{1}, \\ \\ E_{2}{\circ_{1,Z}}E_{6} = E_{1}, \; \; \; \;
E_{5}{\circ_{1,Z}}E_{6} = E_{3}. \\
\end{array}
\end{equation}

These five cases, covering all $4+6+3+4+4=21$ ordered quaternionic
triples, prove (17). Together with (12) and (18), they prove
%
%
\begin{equation}
\fbox{$ {\bf O}_{1,Z} \simeq {\bf O}.
$}
\end{equation} \\
So ${\bf O}_{1,Z}$ is in fact a copy of the
octonions. \\

\begin{flushleft}
GENERAL  ${\bf O}_{X,Y}$. \\
\end{flushleft}

In general, starting from any copy of the octonions, the ${\bf
O}_{1,Z}$ modification will be another copy of the octonions.  In
particular,
%
%
\begin{equation}
({\bf O}_{X})_{1,Z} \simeq {\bf O}.
\end{equation}
The product of ${\bf O}_{X}$ is
$$
A{\circ_{X}}B = (AX)(X^{\dagger}B).
$$
Now modify this to the product of $({\bf O}_{X})_{1,Z}$:
%
%
\begin{equation}
\begin{array}{cl}
A{\circ_{X}}(Z^{\dagger}{\circ_{X}}B) & =
(AX)[X^{\dagger}((Z^{\dagger}X)(X^{\dagger}B))] \\ \\ & =
(AX)[X^{\dagger}(X((X^{\dagger}Z^{\dagger})B))] \\ \\ & =
(AX)[(X^{\dagger}Z^{\dagger})B] \\ \\ & = (AX)(Y^{\dagger}B), \\ \\
\end{array}
\end{equation}
(see {\bf [1][2][3]) where we define
%
%
\begin{equation}
Y=ZX.
\end{equation}
Therefore, for all $X,Y \in S^{7}$,
%
%
\begin{equation}
\fbox{$
{\bf O}_{X,Y} \simeq {\bf O}.
$}
\end{equation}
Note that by virtue of (27),
$$
YX^{\dagger} = Z
$$
is the identity of both ${\bf O}_{1,Z}$ and ${\bf O}_{X,Y}$. \\

Finally, in {\bf [3]} it was shown how the X-product could be used
to generate all the 480 renumberings of the $e_{a}, \; a=1,...,7$,
which leave $e_{0}=1$ fixed as the identity.  There are 7680
renumberings of the entire collection, $e_{a}, \; a=0,...,7$, and
the XY-product plays exactly the same role in this context.
In addition, in the X-product case the 480 renumberings arose from a
pair of octonion $E_{8}$ lattices.  The XY-product renumberings are
related in a similar fashion to the pair of octonion $\Lambda_{16}$
lattices developed in {\bf [4]}.  I'll leave it to the reader to prove
this, or the reader can wait for a complete development in the
monograph which is in preparation.

\end{document}